\documentstyle[twoside,fleqn,espcrc2,epsf]{article}

\def\simgt{\rlap{\lower 3.5 pt\hbox{$\mathchar \sim$}}\raise 1pt \hbox {$>$}}
\def\simlt{\rlap{\lower 3.5 pt\hbox{$\mathchar \sim$}}\raise 1pt \hbox {$<$}}

\title{Continuum Limit of the Heavy-Light Decay
       Constant with the Quenched Wilson Quark
       Action\thanks{presented by S. Hashimoto}}

\author{JLQCD Collaboration \\[2mm]
        S.~Aoki\address{Institute of Physics, University of
        Tsukuba, Tsukuba, Ibaraki 305, Japan},
        M.~Fukugita\address{Yukawa Institute for Theoretical Physics,
        Kyoto University, Kyoto 606, Japan},
        S.~Hashimoto\address{National Laboratory for High
        Energy Physics (KEK), Tsukuba, Ibaraki 305, Japan},
        N.~Ishizuka$^{\rm a}$,
        Y.~Iwasaki$^{\rm a,}$\address{Center for Computational Physics,
        University of Tsukuba, Tsukuba, Ibaraki 305, Japan},
        K.~Kanaya$^{\rm a,d}$,
        Y.~Kuramashi$^{\rm c}$,
        H.~Mino\address{Faculty of Engineering, Yamanashi
        University, Kofu 400, Japan},
        M.~Okawa$^{\rm c}$, A.~Ukawa$^{\rm a}$,
        T.~Yoshi\'e$^{\rm a,d}$ }

\begin{document}

\begin{abstract}
We explore the problems in calculating the decay constant
for heavy-light mesons using the quenched Wilson quark
action for both heavy and light quarks. 
We find that the continuum limit of the decay constant
is reasonably converged among different prescriptions
after the continuum limit is taken.
A number of technical problems associated with prescriptions
are also addressed.
\end{abstract}

\maketitle

\section{Introduction}
Extrapolation to the continuum limit is an indispensable
procedure to obtain physics from lattice calculations.
For the system including heavy quarks ($Q$) this extrapolation
might no longer be straightforward,
since $O(m_{Q}a)$ errors can uncontrollably be large.
In this work we numerically explore the problem of the
continuum limit for the decay constant of heavy-light mesons 
using the Wilson heavy quark action in the quenched
approximation. 

\section{Simulation}
Table~\ref{table:simulation_parameters} summarizes
parameters we used in our simulations carried out on
VPP-500/80 at KEK.
The improvement we have made since
Lattice 95\cite{JLQCD_95} is the adoption of a smeared
source for heavy quarks to suppress contaminations from
excited states, and an addition of a run at $\beta=5.9$.
We use as a smearing function the wave function of
heavy-light mesons as measured in the simulation.
The size of lattice is chosen so that physical volume is
roughly constant against $\beta$. 
We take the physical scale of the lattice estimated from
the charmonium spin-averaged 1S-1P mass splitting. 

\begin{table}
\setlength{\tabcolsep}{0.2pc}
\caption{Simulation Parameters}
\vspace{-3mm}
\label{table:simulation_parameters}
\begin{center}
\begin{tabular}{llll} \hline
  $\beta$         & 5.9 & 6.1 & 6.3 \\
  \hline
  size            & $\mbox{16}^{3}\times\mbox{40}$ &
                    $\mbox{24}^{3}\times\mbox{64}$ &
                    $\mbox{32}^{3}\times\mbox{80}$  \\
  \#conf.         & 150        & 100        & 100  \\
  $a_{1S-1P}^{-1}$(GeV)
                  & 1.84(03)   & 2.54(07)   & 3.36(11) \\
  \hline
\end{tabular}
\end{center}
\vspace*{-10mm}
\end{table}

\section{Kinetic versus pole mass}
The effective Hamiltonian for Wilson quark with a large mass $m_Qa$ 
takes the form\cite{El-Khadra_Kronfeld_Mackenzie_96},
\begin{equation}
   \label{eq:Hamiltonian}
  H = \bar{\Psi} \left[ m_{pole}
          - \frac{\mbox{\boldmath D}^{2}}{2m_{kin}}
          - \frac{\mbox{\boldmath $\sigma$} \cdot
                  \mbox{\boldmath B}}{2m_{mag}}+ \cdots
          \right] \Psi,
\end{equation}
where all mass parameters may differ by $O(m_{Q}a)$.

\begin{figure}[tb]
\centerline{\epsfxsize=7.0cm \epsffile{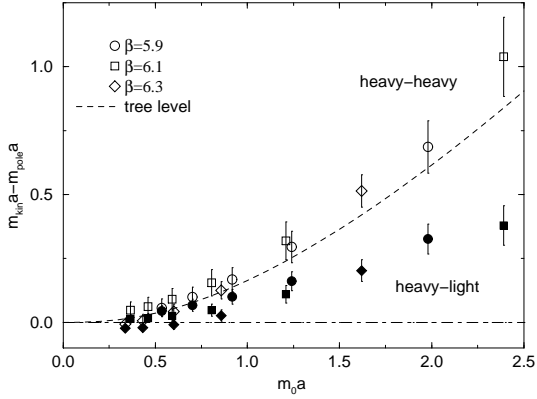}}
\vspace*{-0.9cm}
\caption{
 $m_{kin}a-m_{pole}a$ for heavy-light (filled symbols) and 
 $(m_{kin}a-m_{pole}a)/2$ for heavy-heavy (open symbols) mesons.
 Dashed curve represents tree level value for free Wilson quark.
}
\label{fig:mkin-mpole}
\vspace*{-8mm}
\end{figure}

In Figure~\ref{fig:mkin-mpole} we plot $m_{kin}a-m_{pole}a$
as a function of the bare quark mass for heavy-light
and heavy-heavy mesons, where $m_{kin}a$ is
extracted from measured dispersion relations. 
We expect the two results to agree, since the 
binding energy should cancel in the mass difference.
The data, however, show that they do not, implying that we
cannot determine the kinetic mass consistently from
heavy-light and heavy-heavy systems\cite{SCRI_95}. 
Even worse, as $\beta$ increases, $m_{kin}a-m_{pole}a$
undershoots and becomes negative in a region 
$m_{Q}a \simlt 0.7$. 

\begin{figure}[tb]
\centerline{\epsfxsize=7.0cm \epsffile{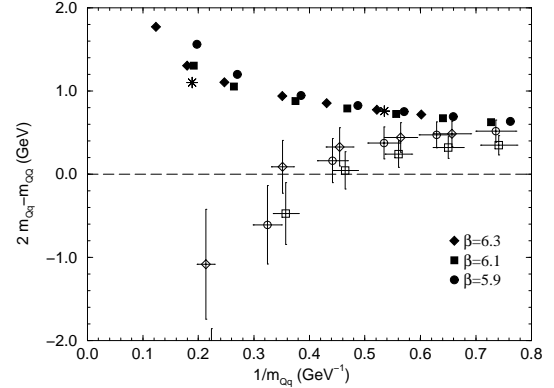}}
\vspace*{-0.9cm}
\caption{$2m_{Q\bar{q}}-m_{Q\bar{Q}}$ calculated with 
  pole mass (filled symbols) and kinetic
  mass (open symbols) as compared with experiment(asterisks).} 
\label{fig:2mQq-mQQ}
\vspace*{-8mm}
\end{figure}

We plot in Figure~\ref{fig:2mQq-mQQ}
$2m_{Q\bar{q}}-m_{Q\bar{Q}}$, with $m_{Q\bar{q}}$ mass of a 
heavy-light meson and $m_{Q\bar{Q}}$ for a heavy-heavy
meson.  This mass difference as obtained with the two mass definitions 
is compared with experimental values (asterisks).
The pole mass gives values close to the
experiment, while the results with the kinetic mass
definition stays far away from the experiment.
Whereas we naively expect that $m_{kin}$ is the most
adequate mass parameter of the system, as the kinetic energy
term dominantly controls the dynamics of the system, our
results show that the kinetic mass measured from simulations
suffers from the effects arising from higher
order terms of $1/m_{Q}$ which receive large finite
lattice corrections\cite{Kronfeld_96}. 
For this reason we cannot conclude that the use of the kinetic mass is 
superior to the pole mass.
 
\section{Heavy-Light Decay Constant}

We study the problem of the continuum limit by taking a
number of alternative prescriptions for the meson mass
parameter and the wave function normalization.
(IA) The simplest choice is use of the pole mass and the
naive normalization $(2K_{Q})^{1/2}$. 
(IB) Alternatively, we may take 
$(8K_{c}-6K_{Q})^{1/2}=(2K_{Q})^{1/2}e^{m_{pole}^{(quark)}a/2}$ 
as the wave function normalization (referred to as the KLM
normalization).
(IIA) The next choice is use of the kinetic mass obtained
from simulations together with the KLM normalization.
In this prescription $O(m_{Q}a)$ corrections are removed
from leading terms of $1/m_{Q}$.
(IIB) In order to avoid the pathology we have seen with the
kinetic mass, we may consider yet another definition of the
`kinetic mass' estimated from the pole mass using the tree
level relation\cite{Bernard_Labrenz_Soni_94},
$m_{kin}^{(meson)}=m_{pole}^{(meson)}
+(m_{kin}^{(quark)}-m_{pole}^{(quark)})$, 
while keeping the KLM normalization.
The tadpole improved perturbative $Z$-factor is used
throughout our analysis.

\begin{figure}[tb]
\centerline{\epsfxsize=7.0cm \epsffile{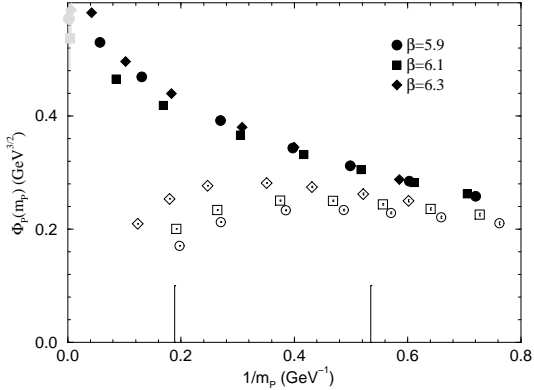}}
\vspace*{-1cm}
\caption{$\Phi_{P}(m_{P})$ versus $1/m_{P}$ with the prescription
  IA (open symbols) and with IIB (filled symbols).
  Results with the static approximation are also shown at
  $1/m_{P}$=0.}
\label{fig:fsqrtm}
\vspace*{-10mm}
\end{figure}

Figure~\ref{fig:fsqrtm} shows the scale invariant quantity
$  \Phi_{P}(m_{P})
  =[\alpha_s(m_{P})/
    \alpha_s(m_{B})]^{2/\beta_{0}}
   f_{P}\sqrt{m_{P}} $
as a function of $1/m_{P}$ for cases (IA) and (IIB).
The naive normalization gives significantly smaller values
than the KLM normalization does for heavy quarks; as is well known 
$f_{P}\sqrt{m_{P}}$ even decreases towards a large quark
mass. 
This contrasts to case (IIB), in which it increases with the
meson mass and is smoothly extrapolated to the value
estimated in the static limit\cite{FNAL_95}.
The method (IB) also approaches the static limit, the
difference from (IIB) being an $O(e^{-m_{P}})$ finite mass
correction rather than
$O(1/m_{P})$\cite{Bernard_Labrenz_Soni_94}. 

\begin{figure}[tb]
\centerline{\epsfxsize=7.0cm \epsffile{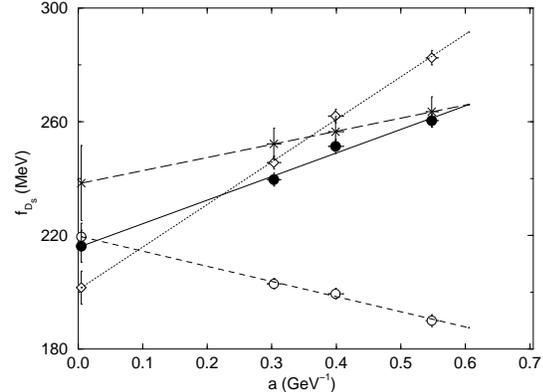}}
\vspace*{-1cm}
\caption{Continuum extrapolation of $f_{D_{s}}$ with four
  methods: IA (open circles), IB (open diamonds), IIA (crosses) and
  IIB (filled circles)}
\label{fig:fDs_continuum}
\vspace*{-7mm}
\end{figure}

Extrapolation to the continuum limit is illustrated in
Figure~\ref{fig:fDs_continuum} for $f_{D_{s}}$
assuming linear $O(m_{Q}a)$ behavior.
Roughly speaking, all prescriptions give a reasonably
convergent answer, and the variation from one to
another is an $O((m_{Q}a)^{2})$ effect.
At a more precise look, however, the result for (IIA) is
sizably deviated from the others.
This is apparently caused by a pathological behavior of
$m_{kin}a-m_{pole}a$ in the charm mass region as discussed 
above. 
We note that the simplest prescription (IA) gives a result
in agreement with the others in spite of a large
extrapolation. The difference from (IB) by an exponential 
factor $e^{m_{pole}^{(quark)}a/2}$ in the KLM 
normalization does not invalidate the linear extrapolation 
for this case since $O((m_Qa)^2)$ terms are small for c-quark.

\begin{figure}[tb]
\centerline{\epsfxsize=7.0cm \epsffile{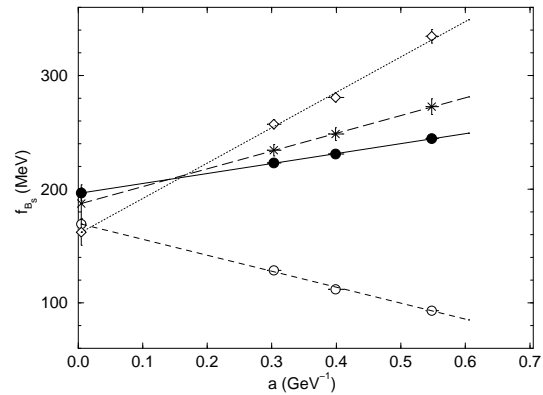}}
\vspace*{-0.8cm}
\caption{Same as Figure~\protect\ref{fig:fDs_continuum} for $f_{B_{s}}$.}
\label{fig:fBs_continuum}
\vspace*{-8mm}
\end{figure}

The results for $f_{B_{s}}$ are also well convergent
in the continuum limit (Figure~\ref{fig:fBs_continuum}).
Although the remaining systematic uncertainty cannot be 
properly estimated for the prescription using the pole mass
for $m_{Q}a$ is larger than unity for the $b$-quark, we 
expect that the remaining errors are reasonably small for
(IIB).

After continuum extrapolation we obtain
\begin{tabular}{ll}
$f_{D}$     = 202(8)$^{+24}_{-11}$ MeV &
$f_{D_{s}}$ = 216(6)$^{+22}_{-15}$ MeV
\end{tabular}
\begin{tabular}{ll}
$f_{B}$     = 179(11)$^{+2}_{-31}$ MeV &
$f_{B_{s}}$ = 197(7)$^{+0}_{-35}$ MeV
\end{tabular}
where we take the value with prescription (IIB) for our
central values. 
The first error is statistical and the second represents a
spread over four prescriptions, which gives an estimate of
$O((m_{Q}a)^{2})$. 
Finally we emphasize that convergence of different extrapolations to the 
continuum limit both from above and below bolsters the reliability of 
our results, at least for charmed mesons.  
We expect that the errors for bottom mesons are at most within the quoted 
errors.

\end{document}